\pdfoutput=1
\documentclass[sigplan,nonacm]{acmart}
\makeatletter
\def\@ACM@checkaffil{
    \if@ACM@instpresent\else
    \ClassWarningNoLine{\@classname}{No institution present for an affiliation}%
    \fi
    \if@ACM@citypresent\else
    \ClassWarningNoLine{\@classname}{No city present for an affiliation}%
    \fi
    \if@ACM@countrypresent\else
        \ClassWarningNoLine{\@classname}{No country present for an affiliation}%
    \fi
}
\makeatother

\setcopyright{none}

\usepackage{stmaryrd}

\usepackage{amsmath}

\usepackage{textcomp}

\usepackage[pdf]{graphviz}
\usepackage{mathrsfs}

\DeclareMathAlphabet{\mathcal}{OMS}{cmsy}{m}{n}
\usepackage{cancel}

\usepackage{centernot}

\usepackage{pgfplots, pgfplotstable}
\pgfplotsset{compat=1.7}
\usepgfplotslibrary{fillbetween}
\usetikzlibrary{patterns}
\pgfmathdeclarefunction{gauss}{2}{\pgfmathparse{1/(#2*sqrt(2*pi))*exp(-((x-#1)^2)/(2*#2^2))}}
\pgfmathdeclarefunction{nil}{1}{\pgfmathparse{0.001}}

\usepackage{arydshln}
\usepackage{adjustbox}
\usepackage{enumerate}
\usepackage{tikz-cd}
\usetikzlibrary{calc}
\usepackage{amsfonts}
\usepackage{bussproofs}

\usepackage{float}

\usepackage{tikz-3dplot}
\usetikzlibrary{3d}
\usetikzlibrary{calligraphy}
\newif\ifshowcellnumber
\showcellnumbertrue

\usepackage{algorithm}
\usepackage{algpseudocode}
\usepackage{algorithmicx}
\usepackage{sourcecodepro}
\usepackage{tikz-qtree}
\usepackage{amsthm}
\usepackage{bm}
\usetikzlibrary{bayesnet}
\usetikzlibrary{arrows}
\usepackage{subcaption}
\usetikzlibrary{backgrounds}
\usetikzlibrary{tikzmark}

\usepackage{mathtools}
\usepackage{tikz}
\usepackage{listings}

\usepackage[skins,breakable,listings]{tcolorbox}

\tcbset{
  enhanced jigsaw,
  breakable,
  listing only,
  boxsep=-1pt,
  top=-1pt,
  bottom=-0.5pt,
  right=-0.5pt,
  overlay first={
    \node[black!50] (S) at (frame.south) {\Large\ding{34}};
    \draw[dashed,black!50] (frame.south west) -- (S) -- (frame.south east);
  },
  overlay middle={
    \node[black!50] (S) at (frame.south) {\Large\ding{34}};
    \draw[dashed,black!50] (frame.south west) -- (S) -- (frame.south east);
    \node[black!50] (S) at (frame.north) {\Large\ding{34}};
    \draw[dashed,black!50] (frame.north west) -- (S) -- (frame.north east);
  },
  overlay last={
    \node[black!50] (S) at (frame.north) {\Large\ding{34}};
    \draw[dashed,black!50] (frame.north west) -- (S) -- (frame.north east);
  },
  before={\par\vspace{5pt}},
  after={\par\vspace{\parskip}\noindent}
}

\definecolor{slightgray}{rgb}{0.90, 0.90, 0.90}

\usepackage{soul}
\makeatletter
\def\SOUL@hlpreamble{%
  \setul{}{3.0ex}%
  \let\SOUL@stcolor\SOUL@hlcolor%
  \SOUL@stpreamble%
}
\makeatother

\definecolor{A}{RGB}{6,150,104}
\definecolor{B}{RGB}{196,74,137}
\definecolor{C}{RGB}{117,237,133}
\definecolor{D}{RGB}{246,46,243}
\definecolor{E}{RGB}{89,162,12}
\definecolor{F}{RGB}{113,12,158}
\definecolor{G}{RGB}{191,205,142}
\definecolor{H}{RGB}{51,58,158}
\definecolor{I}{RGB}{244,212,3}
\definecolor{J}{RGB}{37,36,249}
\definecolor{K}{RGB}{253,165,71}
\definecolor{L}{RGB}{27,81,29}
\colorlet{LA}{A!30}
\colorlet{LB}{B!30}
\colorlet{LC}{C!30}
\colorlet{LD}{D!30}
\colorlet{LE}{E!30}
\colorlet{LF}{F!30}
\colorlet{LG}{G!30}
\colorlet{LH}{H!30}
\colorlet{LI}{I!30}
\colorlet{LJ}{J!30}
\colorlet{LK}{K!30}
\colorlet{LL}{L!30}

\colorlet{lred}{red!30}
\colorlet{lorange}{orange!30}
\colorlet{lgreen}{green!30}
\colorlet{lgray}{black!15}
\colorlet{dgray}{black!75}

\usepackage{url}
\usepackage{qtree}

\usepackage{filecontents}
\usepackage{pstricks-add}
\usepackage{alltt}
\usepackage{nicematrix}
\usepackage{graphicx}
\usepackage{ulem}
\usepackage{upquote}
\tikzstyle{every picture}+=[remember picture]
\usepackage{menukeys}

\makeatletter
\DeclareRobustCommand{\cev}[1]{%
    {\mathpalette\do@cev{#1}}%
}
\newcommand{\do@cev}[2]{%
  \vbox{\offinterlineskip
  \sbox\z@{$\m@th#1 x$}%
  \ialign{##\cr
  \hidewidth\reflectbox{$\m@th#1\vec{}\mkern4mu$}\hidewidth\cr
  \noalign{\kern-\ht\z@}
    $\m@th#1#2$\cr
  }%
  }%
}
\makeatother

\makeatletter
\DeclareRobustCommand{\pder}[1]{%
  \@ifnextchar\bgroup{\@pder{#1}}{\@pder{}{#1}}}
\newcommand{\@pder}[2]{\frac{\partial#1}{\partial#2}}
\makeatother

\makeatletter
\def\squiggly{\bgroup \markoverwith{\textcolor{red}{\lower3\p@\hbox{\sixly \char58}}}\ULon}
\makeatother

\usepgfplotslibrary{fillbetween}
\usetikzlibrary{patterns}

\usepackage[symbol,hang,flushmargin]{footmisc}

\newcommand{\mkTrellis}[1]{
  \begin{tikzpicture}
  \def\dx{20pt}
  \def\dy{30pt}
  \newcounter{i}
  \stepcounter{i}
  \node[circle, draw, fill=black!30] (\arabic{i}) at (0,0){};
  \foreach [count=\i] \x in {2,...,#1}{
    \pgfmathsetmacro{\lox}{\x-1}%
    \pgfmathsetmacro{\loxt}{\x-3}%
    \foreach [count=\j] \xx in {-\lox,-\loxt,...,\lox}{
      \pgfmathsetmacro{\jj}{\j-1}%
      \stepcounter{i}
      \pgfmathsetmacro{\kk}{\xx-2}%
      \pgfmathsetmacro{\lbl}{\lox!/(\jj!*(\lox-\jj)!)}
      \ifnum\x<\kk
      \pgfmath\node[circle, draw]  (\arabic{i}) at (\xx*\dx, -\lox*\dy) {};
      \else
      \pgfmath\node[circle, draw, fill=black!30]  (\arabic{i}) at (\xx*\dx, -\lox*\dy) {};
      \fi
    }
  }
  \newcounter{z}
  \newcounter{xn}
  \newcounter{xnn}
  \pgfmathsetmacro{\maxx}{#1 - 1}
  \foreach \x in {1,...,\maxx}{
    \foreach \xx in {1,...,\x}{
      \stepcounter{z}
      \setcounter{xn}{\arabic{z}}
      \addtocounter{xn}{\x}
      \setcounter{xnn}{\arabic{xn}}
      \stepcounter{xnn}
      \draw [<-] (\arabic{z}) -- (\arabic{xn});
      \draw [<-] (\arabic{z}) -- (\arabic{xnn});
    }
  }
  \end{tikzpicture}
}

\newcommand{\dx}{20pt}
\newcommand{\dy}{30pt}
\newcounter{i}
\newcounter{z}
\newcounter{xn}
\newcounter{xnn}
\newcommand{\mkTrellisAppend}[1]{
  \begin{tikzpicture}
  \setcounter{i}{0}
  \setcounter{z}{0}
  \setcounter{xn}{0}
  \setcounter{xnn}{0}
  \stepcounter{i}
  \node[circle, draw] (\arabic{i}) at (0,0){};
  \foreach [count=\i] \x in {2,...,#1}{
    \pgfmathsetmacro{\lox}{\x-1}%
    \pgfmathsetmacro{\loxt}{\x-3}%
    \foreach [count=\j] \xx in {-\lox,-\loxt,...,\lox}{
      \pgfmathsetmacro{\jj}{\j-1}%
      \stepcounter{i}
      \pgfmathsetmacro{\kk}{\xx+2}%
      \pgfmathsetmacro{\lbl}{\lox!/(\jj!*(\lox-\jj)!)}
      \ifnum\x>\kk
      \pgfmath\node[circle, draw, fill=black!30]  (\arabic{i}) at (\xx*\dx, -\lox*\dy) {};
      \else
      \pgfmath\node[circle, draw]  (\arabic{i}) at (\xx*\dx, -\lox*\dy) {};
      \fi
    }
  }
  \pgfmathsetmacro{\maxx}{#1 - 1}
  \foreach \x in {1,...,\maxx}{
    \foreach \xx in {1,...,\x}{
      \stepcounter{z}
      \setcounter{xn}{\arabic{z}}
      \addtocounter{xn}{\x}
      \setcounter{xnn}{\arabic{xn}}
      \stepcounter{xnn}
      \draw [<-] (\arabic{z}) -- (\arabic{xn});
      \draw [<-] (\arabic{z}) -- (\arabic{xnn});
    }
  }
  \end{tikzpicture}
}

\newcommand{\mkTrellisInsert}[1]{
  \begin{tikzpicture}
  \setcounter{i}{0}
  \setcounter{z}{0}
  \setcounter{xn}{0}
  \setcounter{xnn}{0}
  \stepcounter{i}
  \node[circle, draw] (\arabic{i}) at (0,0){};
  \foreach [count=\i] \x in {2,...,#1}{
    \pgfmathsetmacro{\lox}{\x-1}%
    \pgfmathsetmacro{\loxt}{\x-3}%
    \foreach [count=\j] \xx in {-\lox,-\loxt,...,\lox}{
      \pgfmathsetmacro{\jj}{\j-1}%
      \stepcounter{i}
      \pgfmathsetmacro{\mp}{\xx+#1}%
      \pgfmathsetmacro{\mq}{\xx+\x}%
      \pgfmathsetmacro{\lbl}{\lox!/(\jj!*(\lox-\jj)!)}
      \ifnum\x>\mp
      \pgfmath\node[circle, draw, fill=black!30]  (\arabic{i}) at (\xx*\dx, -\lox*\dy) {};
      \else
      \ifnum#1<\mq
      \pgfmath\node[circle, draw, fill=black!30]  (\arabic{i}) at (\xx*\dx, -\lox*\dy) {};
      \else
      \pgfmath\node[circle, draw]  (\arabic{i}) at (\xx*\dx, -\lox*\dy) {};
      \fi
      \fi

    }
  }
  \pgfmathsetmacro{\maxx}{#1 - 1}
  \foreach \x in {1,...,\maxx}{
    \foreach \xx in {1,...,\x}{
      \stepcounter{z}
      \setcounter{xn}{\arabic{z}}
      \addtocounter{xn}{\x}
      \setcounter{xnn}{\arabic{xn}}
      \stepcounter{xnn}
      \draw [<-] (\arabic{z}) -- (\arabic{xn});
      \draw [<-] (\arabic{z}) -- (\arabic{xnn});
    }
  }
  \end{tikzpicture}
}

\usetikzlibrary{automata, positioning, arrows}
\tikzset{
  node distance=3cm, 
  initial text=$ $, 
}

\tikzset{
  treenode/.style = {shape=rectangle, rounded corners,
  draw, align=center,
  top color=white, bottom color=blue!20},
  root/.style     = {treenode, font=\tiny, bottom color=red!30},
  env/.style      = {treenode, font=\tiny},
  dummy/.style    = {circle,draw}
}

\usepackage{stackengine}
\begin{document}

  \title{A Tree Sampler for Bounded Context-Free Languages}
  \begin{abstract}
    In the following paper, we present a simple method for sampling trees with or without replacement from BCFLs. A BCFL is a context-free language (CFL) corresponding to an incomplete string with holes, which can be completed by valid terminals. To solve this problem, we introduce an algebraic datatype that compactly represents candidate parse forests for porous strings. Once constructed, sampling trees is a straightforward matter of sampling integers uniformly without replacement, then lazily decoding them into trees.
  \end{abstract}

  \author{Breandan Considine}
  \affiliation{\institution{McGill University}}
  \email{bre@ndan.co}

  \maketitle

  \section{Introduction}

  A CFG is a quadruple consisting of terminals $(\Sigma)$, nonterminals $(V)$, productions $(P\colon V \rightarrow (V \mid \Sigma)^*)$, and a start symbol, $(S)$. It is a well-known fact that every CFG is reducible to \textit{Chomsky Normal Form}, $P'\colon V \rightarrow (V^2 \mid \Sigma)$, in which every production takes one of two forms, either $w \rightarrow xz$, or $w \rightarrow t$, where $w, x, z: V$ and $t: \Sigma$. For example, the CFG, $P=\{S \rightarrow S S \mid ( S ) \mid ()\}$, corresponds to the CNF:\vspace{-3pt}

  \begin{table}[H]
    \begin{tabular}{llll}
      $P'=\big\{\;S\rightarrow QR \mid SS \mid LR,$ & $L \rightarrow (,$ & $R \rightarrow ),$ & $Q\rightarrow LS\;\big\}$
    \end{tabular}
  \end{table}\vspace{-8pt}

  \noindent Given a CFG, $\mathcal{G}' : \langle \Sigma, V, P, S\rangle$ in CNF, we can construct a recognizer $R: \mathcal{G}' \rightarrow \Sigma^n \rightarrow \mathbb{B}$ for strings $\sigma: \Sigma^n$ as follows. Let $2^V$ be our domain, $0$ be $\varnothing$, $\oplus$ be $\cup$, and $\otimes$ be defined as:\vspace{-10pt}

  \begin{align}
    X \otimes Z = \big\{\;w \mid \langle x, z\rangle \in X \times Z, (w\rightarrow xz) \in P\;\big\}
  \end{align}

  \noindent If we define $\hat\sigma_r = \{w \mid (w \rightarrow \sigma_r) \in P\}$, then initialize $M^0_{r+1=c}(\mathcal{G}', e) = \;\hat\sigma_r$ and solve for $M = M + M^2$, the fixedpoint $M_\infty$ is fully determined by the superdiagonal entries:\vspace{-10pt}

  \begin{align*}
      M^0=\begin{pNiceMatrix}[nullify-dots,xdots/line-style=loosely dotted]
        \varnothing & \hat\sigma_1   & \varnothing & \Cdots & \varnothing \\
        \Vdots      & \Ddots         & \Ddots      & \Ddots & \Vdots\\
                    &                &             &        & \varnothing\\
                    &                &             &        & \hat\sigma_n \\
        \varnothing & \Cdots         &             &        & \varnothing
      \end{pNiceMatrix} &\Rightarrow M_\infty =
      \begin{pNiceMatrix}[nullify-dots,xdots/line-style=loosely dotted]
        \varnothing & \hat\sigma_1   & \Lambda & \Cdots & \Lambda^*_\sigma\\
        \Vdots      & \Ddots         & \Ddots  & \Ddots & \Vdots\\
                    &                &         &        & \Lambda\\
                    &                &         &        & \hat\sigma_n \\
        \varnothing & \Cdots         &         &        & \varnothing
      \end{pNiceMatrix}
  \end{align*}

  \noindent Once obtained, the proposition $[S \in \Lambda^*_\sigma]$ decides language membership, i.e., $[\sigma \in \mathcal{L}(\mathcal{G})]$. This procedure is essentially the textbook CYK algorithm in a linear algebraic notation~\cite{goodman1999semiring}. We are now ready to define the sampling problem as follows:

  \begin{definition}[Completion]
    Let $\underline\Sigma = \Sigma \cup \{\_\}$, where $\_$ represents a hole. We denote $\sqsubseteq: \Sigma^n \times \underline\Sigma^n$ as the relation $\{\langle\sigma', \sigma\rangle \mid \sigma_i: \Sigma \implies \sigma_i' = \sigma_i\}$ and the set $\{\sigma': \Sigma \mid \sigma' \sqsubseteq \sigma\}$ as $\text{H}(\sigma)$. Given a porous string $\sigma: \underline\Sigma$, we want to sample parse trees generated by $\mathcal{G}$ corresponding to $\sigma': \text{H}(\sigma)\cap\ell$.
  \end{definition}

  $\text{H}(\sigma)\cap\ell$ is often a large-cardinality set, so we want a procedure which samples trees uniformly without replacement, without enumerating the whole set, parsing and shuffling it.

  \pagebreak\section{Method}\label{sec:method}

   We define an algebraic data type $\mathbb{T}_3 = (V \cup \Sigma) \rightharpoonup \mathbb{T}_2$ where $\mathbb{T}_2 = (V \cup \Sigma) \times (\mathbb{N} \rightharpoonup \mathbb{T}_2\times\mathbb{T}_2)$\footnote{Given a $T:\mathbb{T}_2$, we may also refer to $\pi_1(T), \pi_2(T)$ as $\texttt{root}(T)$ and $\texttt{children}(T)$ respectively, where children are pairs of conjoined twins.}. Morally, we can think of $\mathbb{T}_2$ as an implicit set of possible trees sharing the same root, and $\mathbb{T}_3$ as a dictionary of possible $\mathbb{T}_2$ values indexed by possible roots, given by a specific CFG under a finite-length porous string. We construct $\hat\sigma_r = \Lambda(\sigma_r)$ as follows:

\vspace{-10pt}\begin{equation*}
  \begin{footnotesize}
\Lambda(s: \underline\Sigma) \mapsto \begin{cases}
\bigoplus_{s\in \Sigma} \Lambda(s) & \text{if $s$ is a hole,} \vspace{5pt}\\
\big\{\mathbb{T}_2\big(w, \big[\langle\mathbb{T}_2(s), \mathbb{T}_2(\varepsilon)\rangle\big]\big) \mid (w \rightarrow s)\in P\big\} & \text{otherwise.}
\end{cases}
  \end{footnotesize}
\end{equation*}

\noindent This initializes the superdiagonal, enabling us to compute the fixpoint $M_\infty$ by redefining $\oplus, \otimes: \mathbb{T}_3 \times \mathbb{T}_3 \rightarrow \mathbb{T}_3$ as:

\begin{equation*}
  \begin{footnotesize}
  X \oplus Z \mapsto \bigcup_{\mathclap{k\in \pi_1(X \cup Z)}}\Big\{k \Rightarrow \mathbb{T}_2(k, x \cup z) \mid x \in \pi_2(X\circ k), z \in \pi_2(Z\circ k)\Big\}
  \end{footnotesize}
\end{equation*}

\begin{equation*}
  \begin{footnotesize}
  X \otimes Z \mapsto \bigoplus_{\mathclap{(w\rightarrow xz) \in P}}\Big\{\mathbb{T}_2\Big(w, \big[\langle X\circ x, Z\circ z\rangle\big]\Big) \mid x \in \pi_1(X), z \in \pi_1(Z)\Big\}
\end{footnotesize}
\end{equation*}

  These operators group subtrees by their root nonterminal, then aggregate their children. Instead of tracking sets, each $\Lambda$ now becomes a dictionary indexed by the root nonterminal, which can be sampled by obtaining $(\Lambda_\sigma^* \circ S): \mathbb{T}_2$, then recursively choosing twins as we describe in \S~\ref{sec:replacement}, or without replacement via enumeration as described in \S~\ref{sec:pairing}.


\subsection{Sampling trees with replacement}\label{sec:replacement}

Given a probabilistic CFG whose productions indexed by each nonterminal are decorated with a probability vector $\mathbf{p}$ (this may be uniform in the non-probabilistic case), we define a tree sampler $\Gamma: (\mathbb{T}_2 \mid \mathbb{T}_2^2) \rightsquigarrow \mathbb{T}$ which recursively samples children according to a Multinoulli distribution:

\begin{equation*}
  \Gamma(T) \mapsto \begin{cases}
        \Gamma\big(\text{Multi} \big(\texttt{children}(T), \mathbf{p}\big)\big) & \text{ if $T: \mathbb{T}_2$ } \\
        \big\langle \Gamma\big(\pi_1(T)\big), \Gamma\big(\pi_2(T)\big) \big\rangle & \text{ if $T: \mathbb{T}_2\times\mathbb{T}_2$ }
  \end{cases}
\end{equation*}

This is closely related to the generating function for the ordinary Boltzmann sampler from analytic combinatorics,

\begin{equation*}
  \Gamma C(x) \mapsto \begin{cases}
  \text{Bern} \left(\frac{A(x)}{A(x) + B(x)}\right) \rightarrow \Gamma A(x) \mid \Gamma B(x) & \text{ if } \mathcal{C}=\mathcal{A}+\mathcal{B} \\
  \big\langle \Gamma A(x), \Gamma B(x)\big\rangle & \text{ if } \mathcal{C}=\mathcal{A} \times \mathcal{B}
  \end{cases}
\end{equation*}

\noindent however unlike Duchon et al.~\cite{duchon2004boltzmann}, our work does not depend on rejection to guarantee exact-size sampling, as all trees contained in $\mathbb{T}_2$ will necessarily be the same width.

\pagebreak\subsection{Sampling without replacement}\label{sec:pairing}

The type $\mathbb{T}_2$ of all possible trees that can be generated by a CFG in Chomksy Normal Form corresponds to the fixpoints of the following recurrence, which tells us that each $\mathbb{T}_2$ can be a terminal, or a nonterminal and a (possibly empty) sequence of nonterminal pairs and their two children:\vspace{-10pt}

\begin{equation*}
  L(p) = 1 + p L(p) \phantom{addspace} P(a) = \Sigma + V L\big(V^2P(a)^2\big)
\end{equation*}

Given a $\sigma: \underline\Sigma$, we construct $\mathbb{T}_2$ from the bottom-up, and sample from the top-down. Depicted below is a partial $\mathbb{T}_2$, where red nodes are \texttt{root}s and blue nodes are \texttt{children}:

\begin{figure}[H]
\resizebox{0.9\columnwidth}{!}{
  \begin{tikzpicture}
  [
    grow                    = right,
    sibling distance        = 3em,
    level distance          = 5em,
    edge from parent/.style = {draw, -latex},
    every node/.style       = {font=\footnotesize},
    sloped
  ]
  \node [root] {S}
  child { node [env] {BC}
    child { node [root] {B}
      child { node [env] {RD}
        child { node [root] {R} }
        child { node [root] {D} }
      }
    }
    child { node [root] {C}
      child { node [env] {$\ldots\vphantom{BB}$} }
    }
  }
  child { node [env] {$\ldots\vphantom{BB}$} }
  child { node [env] {AB}
    child { node [root] {A}
      child {
        node [env] {QC}
        child { node [root] {Q} }
        child { node [root] {C} }
      }
      child { node [env] {$\ldots\vphantom{BB}$} }
    }
    child { node [root] {B}
      child { node [env] {RD}
        child { node [root] {R}  }
        child { node [root] {D}  }
      }
    }
  };
  \end{tikzpicture}
}
\caption{A partial $\mathbb{T}_2$ for the grammar with productions $P=\{S \rightarrow BC \mid \ldots \mid AB, B\rightarrow RD \mid \ldots, A\rightarrow QC \mid \ldots\}$.}
\end{figure}
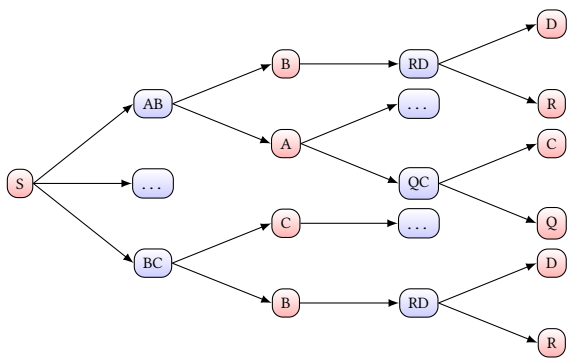

To obtain the total number of trees with breadth $n$, we abstractly parse the porous string using the algebra defined in \S~\ref{sec:method}, and compute the total number of trees using:

\begin{equation*}
  |T: \mathbb{T}_2| \mapsto \begin{cases}
    1  & \text{if $T$ is a leaf,} \\
    \sum_{\langle T_1, T_2\rangle \in \texttt{children}(T)} |T_1| \cdot |T_2| & \text{otherwise.}
  \end{cases}
\end{equation*}

To sample all trees in a given $T: \mathbb{T}_2$ uniformly without replacement, we then construct a modular pairing function $\varphi: \mathbb{T}_2 \rightarrow \mathbb{Z}_{|T|} \rightarrow \texttt{BTree}$, which is defined as follows:

\begin{small}
\begin{equation*}\label{eq:pairing}
\varphi(T: \mathbb{T}_2, i: \mathbb{Z}_{|T|}) \mapsto \begin{cases}
\texttt{BTree}\big(\texttt{root}(T)\big) \text{if $T$ is a leaf,} \vspace{5pt}\\
\textbf{let } F(n) = \sum_{\langle l, r\rangle \in \texttt{children}[0 \ldots n]}|l|\cdot|r|,\\
\phantom{\textbf{let }} F^{-1}(u)=\inf \big\{x \mid u \leq F(x)\big\},\\
\phantom{\textbf{let }} q=i - F\big(F^{-1}(i)\big),\\
\phantom{\textbf{let }} l, r = \texttt{children}[t],\\
\phantom{\textbf{let }} q_1, q_2 =\big\langle\lfloor\frac{q}{|r|}\rfloor, q \pmod{|r|}\big\rangle,\\
\phantom{\textbf{let }} T_1, T_2 = \big\langle\varphi(l, q_1), \varphi(r, q_2)\big\rangle \textbf{ in } \\
\texttt{BTree}\big(\texttt{root}(T), T_1, T_2\big) \text{otherwise.} \\
\end{cases}
\end{equation*}
\end{small}


Then, instead of sampling trees, we can simply sample integers uniformly without replacement from $\mathbb{Z}_{|T|}$ using a full cycle PRNG, and lazily decode them into trees.

%
%

  \section{Prior work}

  Our work is closely related to Boltzmann sampling~\cite{duchon2004boltzmann} in the case of sampling with replacement, but does not use rejection. Piantodosi~\cite{piantadosi2023enumerate} defines a similar construction in the case of sampling without replacement, but it assumes the CFG generates an infinite language and its productions have a certain ordering. In the setting where the template contains only holes, BCFL completion coincides with the Chomsky-Sch\"utzenberger enumeration theorem~\cite{panholzer2005grobner}, which provides a constructive method for counting finite-length words in unambiguous CFLs (i.e., $\ell\cap\Sigma^n$). Our construction is more general, being designed to handle any CFG and template, regardless of ambiguity, finitude, or production ordering.

  Loosely adapted from semiring parsing~\cite{goodman1999semiring} and Valiant's algorithm~\cite{valiant1975general}, our parser also supports bounded generation. We construct a nested datatype~\cite{bird1998nested} that compactly represents candidate parse forests and which can be used to sample trees with or without replacement by sampling a finite integer range, enabling communication-free parallelization. 

  \section{Conclusion}

  We have presented a novel sound and complete algorithm for sampling trees in bounded context-free languages with and without replacement. This technique has applications to code completion and program repair. In future work, we intend to provide a proof of correctness and extend our technique to handle sampling from Boolean and conjunctive languages. A reference implementation for the $\mathbb{T}_2$ datatype is provided in Kotlin and may be found at the \href{https://github.com/breandan/galoisenne/blob/adcb90ac775f17582c5f9fbc4da041b0cf4bf3dc/src/commonMain/kotlin/ai/hypergraph/kaliningraph/parsing/SeqValiant.kt}{URL} linked below.\footnote{\url{https://github.com/breandan/galoisenne/blob/3a2e811e8652ba29891aa21789ef0836ed19d257/src/commonMain/kotlin/ai/hypergraph/kaliningraph/parsing/SeqValiant.kt}}

  \section{Acknowledgements}

  The author wishes to thank David Bieber for mentioning analytic combinatorics during a sunrise hike and David Yu-Tung Hui for sharing his thoughts on Boltzmann sampling.

  \bibliography{acmart}


\begin{thebibliography}{6}


\ifx \showCODEN    \undefined \def \showCODEN     #1{\unskip}     \fi
\ifx \showDOI      \undefined \def \showDOI       #1{#1}\fi
\ifx \showISBNx    \undefined \def \showISBNx     #1{\unskip}     \fi
\ifx \showISBNxiii \undefined \def \showISBNxiii  #1{\unskip}     \fi
\ifx \showISSN     \undefined \def \showISSN      #1{\unskip}     \fi
\ifx \showLCCN     \undefined \def \showLCCN      #1{\unskip}     \fi
\ifx \shownote     \undefined \def \shownote      #1{#1}          \fi
\ifx \showarticletitle \undefined \def \showarticletitle #1{#1}   \fi
\ifx \showURL      \undefined \def \showURL       {\relax}        \fi
\providecommand\bibfield[2]{#2}
\providecommand\bibinfo[2]{#2}
\providecommand\natexlab[1]{#1}
\providecommand\showeprint[2][]{arXiv:#2}

\bibitem[\protect\citeauthoryear{Bird and Meertens}{Bird and Meertens}{1998}]%
        {bird1998nested}
\bibfield{author}{\bibinfo{person}{Richard Bird} {and} \bibinfo{person}{Lambert
  Meertens}.} \bibinfo{year}{1998}\natexlab{}.
\newblock \showarticletitle{Nested datatypes}. In
  \bibinfo{booktitle}{\emph{International Conference on Mathematics of Program
  Construction}}. Springer, \bibinfo{pages}{52--67}.
\newblock
\urldef\tempurl%
\url{http://cs.ox.ac.uk/richard.bird/online/BirdMeertens98Nested.pdf}
\showURL{%
\tempurl}


\bibitem[\protect\citeauthoryear{Duchon, Flajolet, et~al\mbox{.}}{Duchon
  et~al\mbox{.}}{2004}]%
        {duchon2004boltzmann}
\bibfield{author}{\bibinfo{person}{Philippe Duchon}, \bibinfo{person}{Philippe
  Flajolet}, {et~al\mbox{.}}} \bibinfo{year}{2004}\natexlab{}.
\newblock \showarticletitle{Boltzmann samplers for the random generation of
  combinatorial structures}.
\newblock \bibinfo{journal}{\emph{Combinatorics, Probability and Computing}}
  \bibinfo{volume}{13}, \bibinfo{number}{4-5} (\bibinfo{year}{2004}),
  \bibinfo{pages}{577--625}.
\newblock


\bibitem[\protect\citeauthoryear{Goodman}{Goodman}{1999}]%
        {goodman1999semiring}
\bibfield{author}{\bibinfo{person}{Joshua Goodman}.}
  \bibinfo{year}{1999}\natexlab{}.
\newblock \showarticletitle{Semiring parsing}.
\newblock \bibinfo{journal}{\emph{Computational Linguistics}}
  \bibinfo{volume}{25}, \bibinfo{number}{4} (\bibinfo{year}{1999}),
  \bibinfo{pages}{573--606}.
\newblock
\urldef\tempurl%
\url{https://aclanthology.org/J99-4004.pdf}
\showURL{%
\tempurl}


\bibitem[\protect\citeauthoryear{Panholzer}{Panholzer}{2005}]%
        {panholzer2005grobner}
\bibfield{author}{\bibinfo{person}{Alois Panholzer}.}
  \bibinfo{year}{2005}\natexlab{}.
\newblock \showarticletitle{Gr\"{o}bner Bases and the Defining Polynomial of a
  CFG Generating Function}.
\newblock \bibinfo{journal}{\emph{J. Autom. Lang. Comb.}} \bibinfo{volume}{10},
  \bibinfo{number}{1} (\bibinfo{year}{2005}), \bibinfo{pages}{79–97}.
\newblock
\showISSN{1430-189X}


\bibitem[\protect\citeauthoryear{Piantadosi}{Piantadosi}{2023}]%
        {piantadosi2023enumerate}
\bibfield{author}{\bibinfo{person}{Steven~T. Piantadosi}.}
  \bibinfo{year}{2023}\natexlab{}.
\newblock \bibinfo{title}{How to enumerate trees from a context-free grammar}.
\newblock
\newblock
\showeprint[arxiv]{2305.00522}~[cs.CL]


\bibitem[\protect\citeauthoryear{Valiant}{Valiant}{1975}]%
        {valiant1975general}
\bibfield{author}{\bibinfo{person}{Leslie~G Valiant}.}
  \bibinfo{year}{1975}\natexlab{}.
\newblock \showarticletitle{General context-free recognition in less than cubic
  time}.
\newblock \bibinfo{journal}{\emph{Journal of computer and system sciences}}
  \bibinfo{volume}{10}, \bibinfo{number}{2} (\bibinfo{year}{1975}),
  \bibinfo{pages}{308--315}.
\newblock
\urldef\tempurl%
\url{http://people.csail.mit.edu/virgi/6.s078/papers/valiant.pdf}
\showURL{%
\tempurl}


\end{thebibliography}
\end{document}